\documentclass[%
 10pt,
 reprint,
 superscriptaddress,
 amsmath,amssymb,
 aps,
 prl,
]{revtex4-2}

\usepackage{graphicx}
\usepackage{hyperref}

\begin{document}

\title{Hall-Density Collapse as Direct Evidence for Percolative Conduction Onset in Monolayer MoS\texorpdfstring{$_2$}{2} Transistors}

\author{R.~V.~Upadya}
\email{rishabh.xxx@imec.be}
\affiliation{imec, Kapeldreef 75, 3001 Leuven, Belgium} 
\affiliation{Department of Electrical Engineering (ESAT), KU Leuven, 3001 Leuven, Belgium}

\author{M.~Introna}
\affiliation{imec, Kapeldreef 75, 3001 Leuven, Belgium} 
\affiliation{Department of Physics and Astronomy, KU Leuven, 3001 Leuven, Belgium}

\author{X.~Wu}
\affiliation{imec, Kapeldreef 75, 3001 Leuven, Belgium}

\author{C.~J.~L.~de~la~Rosa}
\affiliation{imec, Kapeldreef 75, 3001 Leuven, Belgium}

\author{K.~Banerjee} 
\affiliation{imec, Kapeldreef 75, 3001 Leuven, Belgium} 

\author{D.~Lin} 
\affiliation{imec, Kapeldreef 75, 3001 Leuven, Belgium} 

\author{K.~De~Greve}
\email{kristiaan.degreve@imec.be}
\affiliation{imec, Kapeldreef 75, 3001 Leuven, Belgium} 
\affiliation{Department of Electrical Engineering (ESAT), KU Leuven, 3001 Leuven, Belgium}
\affiliation{Proximus Chair in Quantum Science and Technology, KU Leuven, 3001 Leuven, Belgium}

\begin{abstract}
The origin of the apparent metal-insulator transition in two-dimensional electron systems remains debated. Using temperature-dependent transport and Hall measurements on industry-relevant MoS$_2$ monolayer FETs with amorphous HfO$_2$ dielectrics, we establish classical percolation as the underlying mechanism. We report two complementary signatures of disorder-driven conduction onset: a temperature-independent Hall-density collapse as a function of absolute gate voltage and a vanishing activation energy at a characteristic carrier density $n_{\mathrm{c}}$, identified as the percolation threshold. These findings identify dielectric-interface disorder, rather than channel quality alone, as the key parameter governing device turn-on. The extracted activation barriers reach values substantially exceeding room-temperature thermal energies, indicating that disorder-driven conduction onset remains relevant well beyond cryogenic transport.
\end{abstract}

\maketitle

Two-dimensional (2D) transition metal dichalcogenides (TMDs) are among the leading contenders for transistor technologies beyond silicon because their atomic-scale thickness enables aggressive scaling with reduced short-channel effects \cite{das_transistors_2021, knobloch_improving_2022, lemme_2d_2022}. Their rich spin, valley, and optical degrees of freedom offer opportunities for functionalities beyond conventional electronics \cite{ahn_2d_2020, schaibley_valleytronics_2016, mak_photonics_2016}. Much of the research on 2D TMDs has focused on fundamental transport and electronic phenomena in ultra-clean exfoliated devices, often encapsulated in hBN \cite{wang_electronics_2012, uchiyama_momentum-forbidden_2019}. Real-world applications, however, rely on large-area chemical vapor deposition (CVD) and metal-organic chemical vapor deposition (MOCVD)-grown materials \cite{akinwande_graphene_2019} integrated on technologically relevant substrates \cite{illarionov_process_2024} and embedded within high-$\kappa$ dielectric stacks \cite{xu_scalable_2023}, creating an electrostatic environment that lacks the intrinsic cleanliness of hBN-encapsulated devices and is more susceptible to disorder. Recent studies have suggested that disorder-induced band-tail states may represent a fundamental limitation to transport in realistic two-dimensional semiconductors \cite{han_band_tail_2026}. Understanding how dielectric and interfacial disorder shapes charge transport in this realistic device environment is therefore essential for the realization of TMD electronics \cite{das_sarma_electronic_2011}. To function as effective switches, TMD transistors must simultaneously provide large ON-state currents, low OFF-state leakage, and a sharp, reproducible gate-controlled transition between the ON and OFF states. A central question is therefore not only how high the ON-state current and how low the OFF-state current can be made, but also what physical mechanism governs the onset of conduction in these transistors.

The nature of conduction onset, often framed as an apparent metal-insulator transition (MIT), has remained a key and unresolved question in 2D electron systems for over three decades. The proposed explanations range from an interaction-driven quantum phase transition (QPT) \cite{kravchenko_metalinsulator_2004} to disorder-driven classical percolation \cite{das_sarma_two-dimensional_2005,tracy_observation_2009} and strong-localization/Ioffe-Regel crossover scenarios \cite{das_sarma_two-dimensional_2013,das_sarma_two-dimensional_2014}. Within TMDs, experimental studies have largely favored a percolative interpretation of the onset of conduction, primarily based on conductivity scaling arguments. Percolation scaling has been reported in monolayer MoS$_2$ \cite{moon_anomalous_2019} and multilayer MoSe$_2$ \cite{pradhan_insulator--metal_2023} using carrier densities inferred from geometric gate capacitances, while similar conclusions have been drawn in WSe$_2$ using capacitance-voltage measurements \cite{ali_metalinsulator_2022}. However, these studies primarily rely on conductivity scaling, and the corresponding carrier-density estimates do not directly quantify the mobile carrier population in the presence of disorder and trapping. By contrast, evidence supporting a quantum phase transition has also been reported in multilayer MoS$_2$ \cite{moon_soft_2018}. More recently, studies in ultra-clean and high-mobility TMD platforms have highlighted the continued interplay between interaction-driven and disorder-driven transport, including proposals involving Wigner-crystal physics in ultra-clean devices \cite{huang_electronic_2024,al_mejamai_metal-insulator_2026}. Consequently, despite broad support for a percolative picture, direct experimental verification using independently calibrated mobile-carrier densities remains limited.

Here, we address this issue using industry-relevant monolayer MoS$_2$ field-effect transistors (FETs): MOCVD-grown, transferred using thermal release tape (TRT), and embedded in amorphous HfO$_2$. Using combined Hall and temperature-dependent transport measurements, we identify two complementary signatures of conduction onset: a temperature-independent Hall-density collapse and the vanishing of activation energy at the transition threshold. Together, these measurements provide direct experimental evidence for a percolative conduction mechanism in realistic TMD devices.

\begin{figure*}[htb] 
    \centering 
    \includegraphics[width=\textwidth]{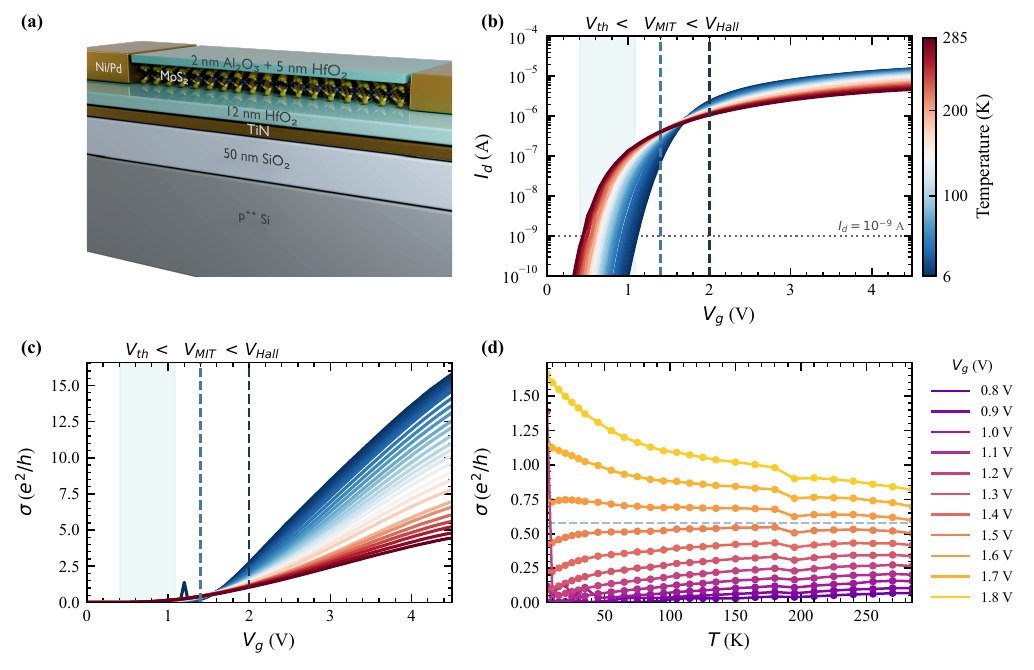} 
    \caption{\textbf{Device architecture and emergence of the three-voltage hierarchy.} (a) Schematic of the monolayer MoS$_2$ FET used in this work. (b) Transfer characteristics measured between 6 and 270 K. (c) Normalized four-terminal conductivity as a function of gate voltage at multiple temperatures. The three characteristic voltage scales are indicated in (b) and (c). From left to right, the shaded region indicates the temperature-dependent range of $V_{\mathrm{th}}$, the light-blue dashed line marks $V_{\mathrm{MIT}}$, and the dark-blue dashed line marks $V_{\mathrm{Hall}}$. (d) Normalized four-terminal conductivity as a function of temperature at fixed gate voltages spanning the insulating and metallic regimes. The dashed horizontal line marks the visual conductivity crossover, which occurs at a slightly higher gate voltage than the final $V_{\mathrm{MIT}}$ extracted from activation-energy analysis. Curves below the crossover exhibit insulating behavior ($d\sigma/dT>0$), whereas curves above the crossover exhibit metallic behavior ($d\sigma/dT<0$).}
    \label{fig:fig1} 
\end{figure*}

Fig.~\ref{fig:fig1}(a) shows the device architecture used throughout this work. Unless otherwise stated, all results presented below are obtained from this device. Four-terminal dc transport and ac Hall measurements were performed at low source-drain bias ($V_{\mathrm{DS}} = 0.1~\mathrm{V}$) in the linear regime to maintain a nearly uniform channel potential and minimize field-induced transport nonuniformities. Fig.~\ref{fig:fig1}(b) shows transfer characteristics measured between 6 and 270 K, revealing a pronounced positive threshold-voltage shift upon cooling. The threshold voltage reported throughout this work is extracted using the constant-current criterion ($I_{\mathrm{d}} = 1~\mathrm{nA}$); threshold voltages obtained from linear extrapolation at the transconductance maximum exhibit similar qualitative trends (see SM, Sec.~S1 \cite{SM}). Such large shifts are consistent with substantial trapping within the dielectric stack and at the MoS$_2$/HfO$_2$ interface, as comparable devices employing hBN dielectrics exhibit markedly reduced threshold-voltage instability~\cite{lan_reliability_2025}.

\begin{figure*}[htb] 
    \centering 
    \includegraphics[width=\textwidth]{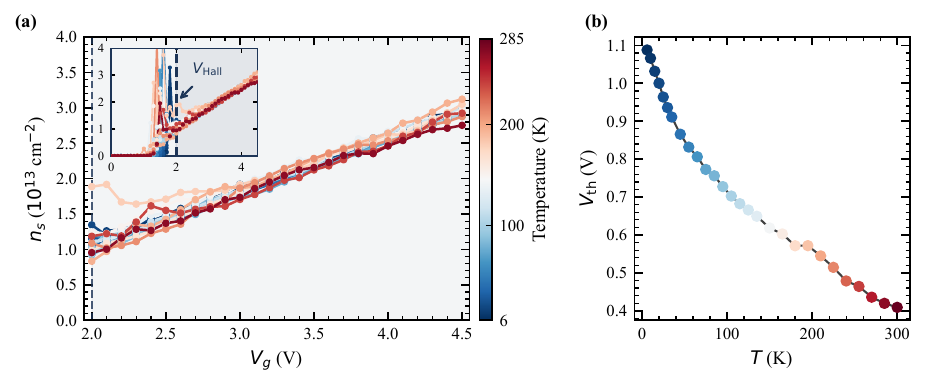} 
    \caption{\textbf{Hall-density collapse and the two-fluid anomaly.} (a) Hall density versus gate voltage measured at 24 temperatures between 6 and 270 K. Inset: full gate-voltage range. Main panel: zoom for $V_{\mathrm{g}} \geq V_{\mathrm{Hall}}$. (b) Threshold voltage $V_{\mathrm{th}}$ extracted using the constant-current criterion ($I_{\mathrm{d}} = 1~\mathrm{nA}$) as a function of temperature.} 
    \label{fig:fig2} 
\end{figure*}

The temperature-dependent transport reveals that a single threshold voltage is insufficient to describe conduction onset in these devices. Instead, three distinct voltage scales emerge naturally from the data, as summarized in Fig.~\ref{fig:fig1}(c). The first is the conventional threshold voltage, $V_{\mathrm{th}}$, widely used in transistor engineering to mark the onset of measurable channel conduction. As a device-level metric, $V_{\mathrm{th}}$ may reflect both channel transport and contact injection. The second is $V_{\mathrm{MIT}}$, defined as the boundary separating thermally activated insulating transport from metallic conduction. Finally, we introduce $V_{\mathrm{Hall}}$, the gate voltage above which Hall measurements yield reliable carrier densities and transport is dominated by extended states. Unlike $V_{\mathrm{th}}$ and $V_{\mathrm{MIT}}$, $V_{\mathrm{Hall}}$ is an operational quantity set by the Hall sensitivity of the measurement. It therefore depends on factors such as carrier mobility and magnetic field. Nonetheless, it remains closely associated with the emergence of a sufficiently large delocalized carrier population. Together, these voltage scales satisfy the ordering $V_{\mathrm{th}} < V_{\mathrm{MIT}} < V_{\mathrm{Hall}}$, demonstrating that measurable current, metallic conduction, and reliable Hall transport occur at distinct stages of conduction onset as carriers progressively delocalize across a disorder-broadened transport landscape.

Fig.~\ref{fig:fig1}(d) reveals a clear transition from insulating to metallic transport through a sign change in $d\sigma/dT$ at fixed gate voltage. The visual crossover occurs near $V_{\mathrm{g}} \approx 1.6~\mathrm{V}$, with conductivity decreasing upon cooling below the transition and increasing upon cooling above it. A more precise determination of $V_{\mathrm{MIT}}$ is obtained from the activation-energy analysis presented later. Notably, the visual conductivity crossover occurs at a slightly higher gate voltage than the final $V_{\mathrm{MIT}}$ extracted from activation-energy analysis (see SM, Sec.~S2 \cite{SM}), consistent with classical percolation theory, where the sign change of $d\sigma/dT$ can occur above the true percolation threshold \cite{das_sarma_two-dimensional_2005}. The separation between $V_{\mathrm{th}}$ and $V_{\mathrm{MIT}}$ therefore reveals an intermediate regime in which measurable current exists despite transport remaining globally insulating. To understand the origin of this hierarchy, we next examine the evolution of the mobile carrier density using Hall measurements.

Fig.~\ref{fig:fig2}(a) shows Hall-measured carrier density $n_{\mathrm{s}}$ as a function of gate voltage $V_{\mathrm{g}}$ at 24 temperatures between 6 and 270 K. Across the full gate-voltage range (inset), reliable Hall densities emerge only above the operational voltage scale $V_{\mathrm{Hall}}$, while the Hall signal becomes increasingly noisy below this point. Strikingly, once $V_{\mathrm{g}} \geq V_{\mathrm{Hall}}$, all curves collapse onto a single trajectory, demonstrating that the mobile carrier density is independent of temperature at fixed absolute gate voltage. By contrast, the threshold voltage extracted from the transfer characteristics shifts by approximately 600--700 mV over the same temperature range, as shown in Fig.~\ref{fig:fig2}(b). 

This observation is unexpected. If transport were governed by a single carrier population, the Hall density would be expected to follow the $V_{\mathrm{th}}$ shift and collapse when plotted against the overdrive voltage ($V_{\mathrm{ov}} = V_{\mathrm{g}} - V_{\mathrm{th}}$). Instead, the collapse occurs against absolute gate voltage ($V_{\mathrm{g}}$). The Hall density therefore remains insensitive to the same temperature-dependent physics responsible for the $V_{\mathrm{th}}$ shift. 

The coexistence of a temperature-independent Hall response and a strongly temperature-dependent $V_{\mathrm{th}}$ is difficult to reconcile within a single-population picture. Instead, it suggests the presence of two distinct carrier populations: a delocalized population detected by Hall measurements and a localized population that contributes to threshold-voltage formation through thermally activated transport. Similar phenomenology has been reported in other strongly disordered electronic systems~\cite{takagi_carrier_2005,kamiya_electronic_2009}, although the microscopic origin of disorder differs substantially. The physical and energetic interpretations of this two-fluid behavior are discussed next. 

\begin{figure*}[htb] 
    \centering 
    \includegraphics[width=\textwidth]{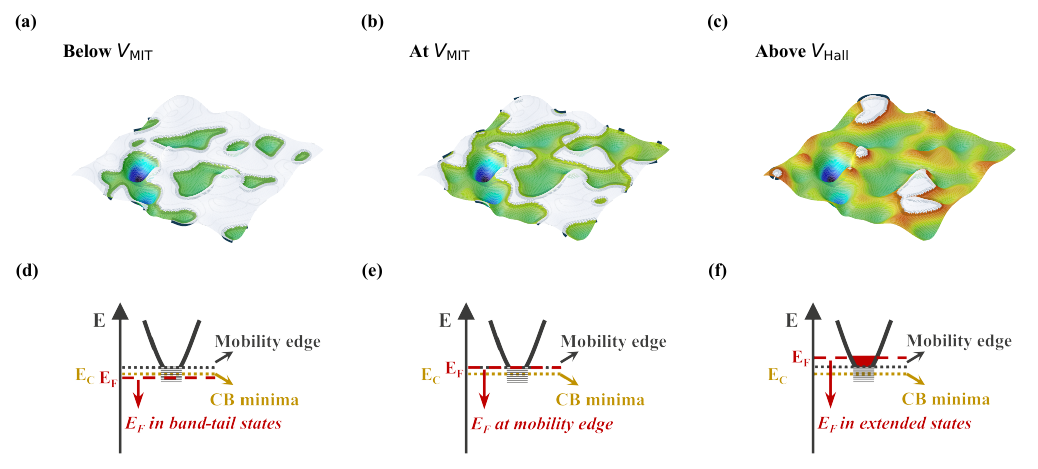} 
    \caption{ \textbf{Conceptual pictures of disorder-driven conduction onset.} (a--c) Real-space evolution of conduction with increasing gate voltage from $V_{\mathrm{th}}$ to $V_{\mathrm{MIT}}$ to $V_{\mathrm{Hall}}$. A fixed disorder potential landscape is assumed throughout; colored regions represent electron-accessible puddles and their progressive delocalization as the Fermi level rises. (d--f) Corresponding energy-space representation showing the Fermi level relative to localized band-tail states, the mobility edge, and extended states.}
    \label{fig:fig3} 
\end{figure*}

Figure 3 presents a conceptual schematic of the disorder-driven percolation framework and is intended to illustrate the physical interpretation of the data rather than a quantitative simulation of the device. In real space [Figs.~\ref{fig:fig3}(a--c)], disorder associated with the dielectric environment, interface states, and material imperfections produces a spatially fluctuating potential landscape, giving rise to electron-rich puddles separated by less conductive regions~\cite{das_sarma_two-dimensional_2005, tracy_observation_2009, klein_impact_2019, shin_characterizing_2023, van_troeye_impact_2025}. The microscopic origin of these fluctuations may involve charged impurities, band-tail disorder, structural defects, or a combination thereof \cite{ducry_oxide_2026}. Regardless of origin, the resulting inhomogeneous potential landscape naturally supports percolative transport. At $V_{\mathrm{th}}$ [Fig.~\ref{fig:fig3}(a)], thermally assisted hopping between isolated puddles yields measurable current, but no continuous conduction path yet spans the device. As $V_{\mathrm{g}}$ increases, the puddles expand and merge until the first system-spanning pathway forms at $V_{\mathrm{MIT}}$ [Fig.~\ref{fig:fig3}(b)], corresponding to the classical percolation threshold. Further gating increases the delocalized carrier population until reliable Hall transport emerges at $V_{\mathrm{Hall}}$ [Fig.~\ref{fig:fig3}(c)].

The same picture can be viewed energetically [Figs.~\ref{fig:fig3}(d--f)]. Disorder broadens the conduction-band edge into a distribution of localized band-tail states, giving rise to an effective mobility edge that separates localized and extended transport. At $V_{\mathrm{th}}$ [Fig.~\ref{fig:fig3}(d)], the Fermi level lies within the band-tail states and conduction occurs through thermally activated hopping. The resulting sensitivity to thermal activation naturally explains the large temperature-dependent shift of $V_{\mathrm{th}}$ observed in Figs.~\ref{fig:fig1}(b) and~\ref{fig:fig2}(b). As the Fermi level rises, it reaches the mobility edge at $V_{\mathrm{MIT}}$ [Fig.~\ref{fig:fig3}(e)], corresponding to the percolation threshold and defining the critical carrier density $n_{\mathrm{c}}$. Further gating moves the Fermi level deeper into the extended-state regime [Fig.~\ref{fig:fig3}(f)], where transport is dominated by delocalized carriers. 

This framework also explains the Hall-density collapse of Fig.~\ref{fig:fig2}. Hall measurements selectively probe carriers occupying extended states above the mobility edge, whereas longitudinal transport remains sensitive to both extended-state conduction and thermally activated transport through localized states. By contrast, $V_{\mathrm{th}}$ is determined much closer to the onset of conduction, where localized carriers dominate the transport response and thermal activation strongly influences the measured current. At higher gate voltages, the delocalized carrier population grows rapidly and dominates transport, while the localized population becomes comparatively less important. Temperature-dependent changes in the localized carrier population therefore produce large shifts in $V_{\mathrm{th}}$, while leaving the Hall density effectively unchanged within experimental uncertainty. The resulting collapse against absolute gate voltage is a direct consequence of the differing roles of localized and delocalized carriers in transport.

\begin{figure*}[htb] 
    \centering 
    \includegraphics[width=\textwidth]{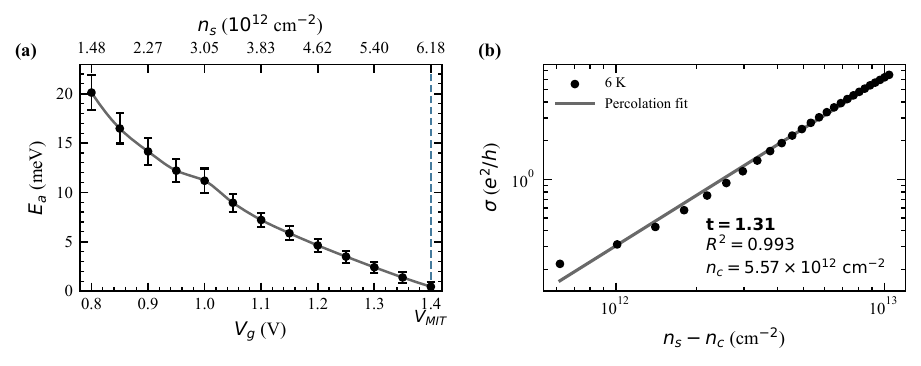} 
    \caption{\textbf{Quantitative validation of the percolation picture.} (a) Activation energy $E_{\mathrm{a}}$ as a function of gate voltage, vanishing at $V_{\mathrm{MIT}}$. (b) Conductivity scaling on the metallic side of the transition, showing $\sigma \propto (n_{\mathrm{s}}-n_{\mathrm{c}})^{t}$ with exponent $t = 1.31$.} 
    \label{fig:fig4} 
\end{figure*}

Fig.~\ref{fig:fig4}(a) shows the activation energy $E_{\mathrm{a}}$ extracted from Arrhenius fits to the insulating-side transport. The activation energy decreases systematically with increasing gate voltage and vanishes within experimental uncertainty at $V_{\mathrm{MIT}} = 1.4~\mathrm{V}$. Within the framework of Fig.~\ref{fig:fig3}, this behavior reflects the progressive reduction of the energy barrier separating the Fermi level from the mobility edge. The point at which $E_{\mathrm{a}}$ vanishes therefore identifies the percolation threshold and defines a critical carrier density $n_{\mathrm{c}} = 6.18 \times 10^{12}~\mathrm{cm}^{-2}$ (see SM, Secs.~S3 and~S4 for the Arrhenius fitting procedure, individual fits, and the carrier-density calibration used to determine $n_{\mathrm{c}}$ \cite{SM}). Importantly, the vanishing of $E_{\mathrm{a}}$ does not require assumptions regarding conductivity scaling or critical exponents. It follows directly from the energetic picture of Fig.~\ref{fig:fig3}, where transport evolves from thermally activated hopping below the mobility edge to extended-state conduction above it. 

An independent test is provided by conductivity scaling on the metallic side of the transition. Fig.~\ref{fig:fig4}(b) shows a fit to the percolation relation 

\begin{equation} 
\sigma \propto (n_{\mathrm{s}}-n_{\mathrm{c}})^t 
\end{equation} 

using the 6 K transport data and fixing $V_{\mathrm{MIT}} = 1.4~\mathrm{V}$ from the independent $E_{\mathrm {a}} \rightarrow 0$ analysis. Excellent agreement is obtained across approximately one decade in $(n_{\mathrm {s}}-n_{\mathrm {c}})$, with an extracted exponent $t=1.31$ and coefficient of determination $R^2=0.993$. The extracted exponent is in close agreement with the theoretical two-dimensional percolation value of $4/3$ ~\cite{stauffer_introduction_2018}. The scaling remains robust well beyond the immediate vicinity of the threshold, supporting a genuine percolative transport law rather than a narrowly local fit near $n_{\mathrm{c}}$. The extracted exponent depends sensitively on the assumed threshold: choosing $V_{\mathrm{MIT}}$ above or below $1.4~\mathrm{V}$ systematically drives the fitted exponent away from the expected percolation value (see SM, Fig.~S6 \cite{SM}). The conductivity scaling therefore provides an independent validation of the transition voltage identified through the $E_{\mathrm a}\rightarrow0$ condition. Although the fitted critical density differs slightly from the Hall-calibrated value, this discrepancy is expected given the sensitivity of conductivity-scaling parameters to fitting window and model assumptions. Additional scaling analyses and device-to-device reproducibility of both the Hall-density collapse and $E_{\mathrm{a}} \rightarrow 0$ signatures are provided in the SM (Secs.~S5 and~S6 \cite{SM}).

Taken together, the Hall-density collapse, vanishing activation energy, and conductivity scaling provide substantially more direct support for percolation than conductivity scaling alone. The temperature-independent Hall response and the identification of a characteristic transition density from independent transport signatures point to a disorder-broadened landscape containing coexisting localized and delocalized carriers. These observations are explained naturally within a disorder-driven percolation framework and do not require invoking a quantum phase transition to account for conduction onset. While interaction-driven physics may become important in cleaner or lower-density systems, particularly in ultra-clean platforms where kinetic energy can be suppressed through moiré or flat-band engineering, it remains an open question whether the apparent metal-insulator transition in technologically relevant TMD devices follows the same interaction-driven physics or is instead dominated by disorder.

While demonstrated here in monolayer MoS$_2$/HfO$_2$ transistors, the physical ingredients underlying the present observations are not unique to this material system. Similar disorder-driven transport phenomena have been observed in silicon MOSFETs, III-V heterostructures, and other disordered two-dimensional carrier systems \cite{das_sarma_two-dimensional_2005, tracy_observation_2009}. The transport hierarchy identified here may therefore provide a more general framework for understanding conduction onset in realistic semiconductor devices. Beyond identifying the percolation threshold through the $E_{\mathrm a}\rightarrow0$ condition, the activation-energy analysis also provides insight into the characteristic energy scale associated with the disorder landscape. The extended activation-energy evolution of the primary device is shown in the SM (Sec.~S3 \cite{SM}), while device-to-device statistics are summarized in Sec.~S7 \cite{SM}. Across the investigated devices, the maximum reliable activation barriers range from approximately $57$--$147~\mathrm{meV}$, substantially exceeding room-temperature thermal energies ($k_{\mathrm B}T\approx26~\mathrm{meV}$ at $300~\mathrm{K}$). These values are of the same order of magnitude as oxide-induced electrostatic potential fluctuations predicted by first-principles calculations of TMD/oxide interfaces~\cite{van_troeye_impact_2025}. This comparison suggests that the disorder landscape responsible for percolative conduction onset remains relevant over a significant fraction of the operating bias range, even at room temperature.

A direct consequence of this picture is the emergence of three distinct voltage scales, $V_{\mathrm{th}}$, $V_{\mathrm{MIT}}$, and $V_{\mathrm{Hall}}$, rather than a single threshold voltage. These voltages correspond to the onset of measurable current, percolative metallic conduction, and reliable Hall transport, respectively, demonstrating that threshold behavior in realistic TMD transistors cannot be fully captured by $V_{\mathrm{th}}$ alone. The associated carrier density $n_{\mathrm{c}}$ therefore emerges as a physically meaningful descriptor of conduction onset and a potentially useful metric for comparing realistic TMD device platforms. In addition, $n_{\mathrm{c}}$ may serve as a useful figure of merit for device and dielectric co-optimization, since it directly reflects the disorder landscape governing low-density transport.

In conclusion, conduction onset in industry-relevant monolayer MoS$_2$ transistors is governed by a disorder-driven percolative mechanism. The resulting framework naturally reconciles the Hall-density collapse, vanishing activation energy, and conductivity scaling observations presented here, while explaining the emergence of the three voltage scales $V_{\mathrm{th}} < V_{\mathrm{MIT}} < V_{\mathrm{Hall}}$. These findings establish a physically grounded picture of threshold behavior in realistic TMD devices. They identify the dielectric environment as a key factor governing conduction onset and reveal disorder-related activation barriers that remain significant even relative to room-temperature thermal energies.

\begin{acknowledgments}
This work was performed at imec and supported through imec's Industrial Affiliation Program on Exploratory Logic and KU Leuven. The authors acknowledge valuable discussions within the ExML team and support from the ADC, LABPDEV, and LABOPS teams for materials growth, device fabrication, characterization, dicing, and bonding.
\end{acknowledgments}


\bibliography{refs}

\end{document}


\title{Supplemental Material for Hall-Density Collapse as Direct Evidence for Percolative Conduction Onset in Monolayer MoS\texorpdfstring{$_2$}{2} Transistors}

\author{R.~V.~Upadya}
\email{rishabh.xxx@imec.be}
\affiliation{imec, Kapeldreef 75, 3001 Leuven, Belgium} 
\affiliation{Department of Electrical Engineering (ESAT), KU Leuven, 3001 Leuven, Belgium}

\author{M.~Introna}
\affiliation{imec, Kapeldreef 75, 3001 Leuven, Belgium} 
\affiliation{Department of Physics and Astronomy, KU Leuven, 3001 Leuven, Belgium}

\author{X.~Wu}
\affiliation{imec, Kapeldreef 75, 3001 Leuven, Belgium}

\author{C.~J.~L.~de~la~Rosa}
\affiliation{imec, Kapeldreef 75, 3001 Leuven, Belgium}

\author{K.~Banerjee} 
\affiliation{imec, Kapeldreef 75, 3001 Leuven, Belgium} 

\author{D.~Lin} 
\affiliation{imec, Kapeldreef 75, 3001 Leuven, Belgium} 

\author{K.~De~Greve}
\email{kristiaan.degreve@imec.be}
\affiliation{imec, Kapeldreef 75, 3001 Leuven, Belgium} 
\affiliation{Department of Electrical Engineering (ESAT), KU Leuven, 3001 Leuven, Belgium}
\affiliation{Proximus Chair in Quantum Science and Technology, KU Leuven, 3001 Leuven, Belgium}

\maketitle 

This Supplemental Material provides additional details supporting the results presented in the main text, including threshold-voltage extraction, identification of the metal-insulator transition, Arrhenius analysis, carrier-density calibration, robustness of the percolation scaling, and reproducibility across multiple devices.

\section{Threshold-voltage extraction} 

The threshold voltage of TMD transistors is not uniquely defined and may depend on the extraction methodology. To verify that the conclusions of the main text are not sensitive to the chosen threshold-voltage definition, threshold voltages were extracted using both a constant-current criterion and linear extrapolation at the transconductance maximum. 

\begin{figure*}[htb]
    \centering 
    \includegraphics[width=\columnwidth]{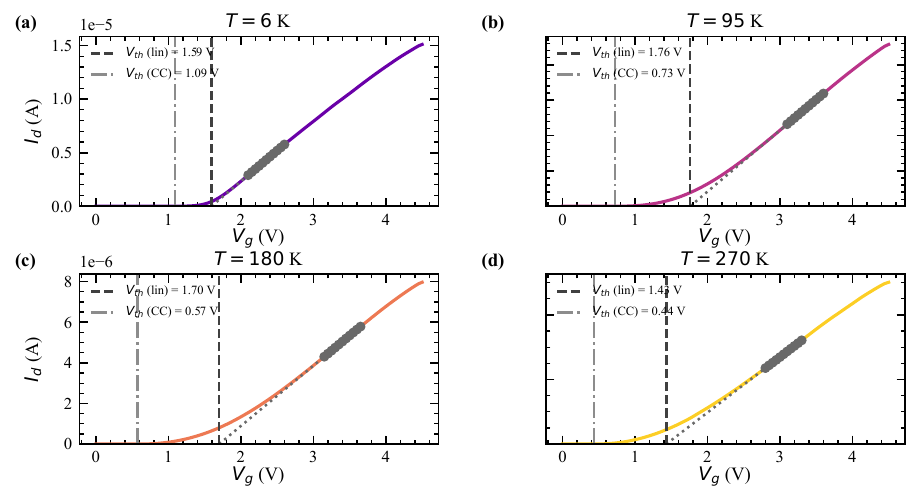} 
    \caption{ \textbf{Comparison of threshold-voltage extraction methods.} 
    Transfer characteristics measured at (a) 6 K, (b) 95 K, (c) 180 K, and (d) 270 K. The threshold voltages obtained from the constant-current criterion ($I_{\mathrm d}=1$ nA) and from linear extrapolation at the transconductance maximum are indicated for each temperature. }
    \label{fig:figS1} 
\end{figure*}

In the main text, the threshold voltage is defined using a constant-current criterion with $I_{\mathrm{d}}=1~\mathrm{nA}$. For each temperature, the corresponding gate voltage was determined directly from the measured transfer characteristic. For comparison, threshold voltages were also extracted using linear extrapolation at the transconductance maximum. In this method, the approximately linear region surrounding the transconductance peak was fitted and extrapolated to zero drain current to obtain the corresponding threshold voltage. Representative examples of both extraction procedures are shown in Fig.~\ref{fig:figS1} for measurements at 6, 95, 180, and 270~K. The constant-current threshold voltage $V_{\mathrm{th,CC}}$ and the threshold voltage obtained from linear extrapolation $V_{\mathrm{th,lin}}$ are indicated for each transfer characteristic.

\begin{figure*}[htb] 
    \centering 
    \includegraphics[width=0.55\columnwidth]{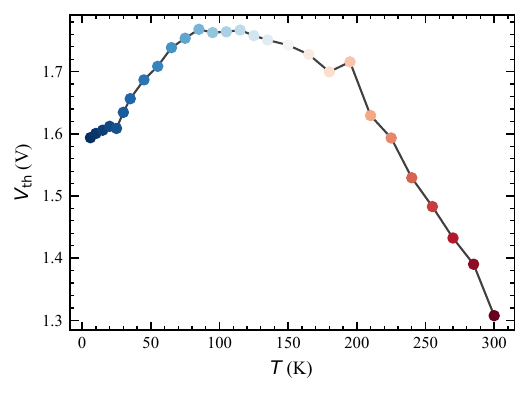}
    \caption{ \textbf{Threshold voltages extracted using linear extrapolation at the transconductance maximum.} 
    Temperature dependence of the threshold voltage obtained using the transconductance-based linear-extrapolation method. }
    \label{fig:figS2} 
\end{figure*}

The temperature dependence of the threshold voltage obtained using the constant-current criterion is shown in Fig.~2(b) of the main text. For comparison, Fig.~\ref{fig:figS2} shows the threshold voltages obtained from linear extrapolation at the transconductance maximum. While the extracted threshold voltages exhibit the same overall increase upon cooling, additional non-monotonic variations are observed at intermediate temperatures.

The transconductance-based threshold depends not only on the position of the transfer characteristic but also on the location of the transconductance maximum, which itself evolves with temperature. As a result, the extracted values include contributions from both effects and can exhibit additional variations beyond the shift of the transfer curve alone. Consequently, the constant-current criterion provides a more direct measure of the threshold-voltage shift and was therefore adopted throughout the main text. The monotonic evolution obtained from the constant-current criterion is also consistent with the systematic shift of the transfer characteristics toward higher gate voltages upon cooling, as shown in Fig.~1(b) of the main text.

Importantly, both extraction methods yield the same qualitative conclusion: the threshold voltage increases substantially upon cooling. The principal conclusions of the main text are therefore independent of the threshold-voltage definition employed.

\clearpage
\section{Visual conductivity crossover versus true \texorpdfstring{$V_{\mathrm{MIT}}$}{VMIT}}

A common experimental approach for identifying the metal-insulator transition is to determine the gate voltage at which the temperature dependence of the conductivity changes sign. In the present device, this visual conductivity crossover occurs between approximately 1.5 and 1.6 V, as shown in Fig.~1(d) of the main text. The same crossover is also visible in the transfer characteristics [Fig.~1(b)] and normalized conductivity-gate-voltage curves [Fig.~1(c)], which provide complementary visualizations of the same underlying transport transition.

From the normalized conductivity curves shown in Fig.~1(d), transport evolves from insulating behavior ($d\sigma/dT > 0$) at lower gate voltages to metallic behavior ($d\sigma/dT < 0$) at higher gate voltages. Based solely on this visual crossover, one would estimate the transition voltage to lie between approximately 1.5 and 1.6 V. 

However, the activation-energy analysis presented in Fig.~4(a) provides a more direct determination of the transition. The activation energy decreases systematically with increasing gate voltage and vanishes within experimental uncertainty at $V_{\mathrm{g}} = 1.4~\mathrm{V}$. Within the disorder-driven percolation framework developed in the main text, this condition identifies the percolation threshold and therefore defines $V_{\mathrm{MIT}}$. 

An independent validation is provided by the conductivity scaling analysis. Using $V_{\mathrm{MIT}} = 1.4~\mathrm{V}$ obtained from the $E_{\mathrm{a}}\rightarrow0$ condition yields a percolation exponent $t = 1.31$, in close agreement with the theoretical two-dimensional percolation value of $4/3$. As shown in Sec.~S5, choosing larger values of $V_{\mathrm{MIT}}$ systematically drives the extracted exponent away from the expected percolation value. The conductivity scaling therefore independently supports $V_{\mathrm{MIT}} = 1.4~\mathrm{V}$ as the physically relevant transition voltage. 

The offset between the visual conductivity crossover and the true percolation threshold is expected within classical percolation theory. Das Sarma \emph{et al.}~\cite{das_sarma_two-dimensional_2005} pointed out that transition densities estimated solely from the sign change of $d\sigma/dT$ are generally expected to exceed the true critical density determined from percolation scaling. Near the percolation threshold, a system-spanning conducting network can form while transport still retains an insulating-like temperature dependence due to the continued influence of localized carriers and thermally activated hopping. Metallic temperature dependence therefore emerges only at higher carrier densities where transport becomes increasingly dominated by extended states. Consequently, the sign change of $d\sigma/dT$ is generally expected to occur above the true percolation threshold. The visual crossover therefore provides only an approximate estimate of the transition, whereas the $E_{\mathrm{a}}\rightarrow0$ condition and conductivity scaling provide more direct probes of the underlying percolation threshold. 

Throughout this work, the metal-insulator transition voltage is therefore defined as $V_{\mathrm{MIT}} = 1.4~\mathrm{V}$, as determined from the vanishing activation energy and independently validated through percolation scaling.

\clearpage
\section{Arrhenius fitting procedure} 

Activation energies were extracted from the temperature dependence of the conductivity on the insulating side of the transition, where transport is well described by thermally activated conduction of the form

\begin{equation} 
    \sigma=\sigma_0 \exp\left( -\frac{E_{\mathrm a}} {k_{\mathrm B}T} \right). 
\end{equation} 

Taking the logarithm yields 

\begin{equation} 
    \ln(\sigma) = \ln(\sigma_0) - \frac{E_{\mathrm a}} {k_{\mathrm B}} \frac{1}{T}, 
\end{equation} 

such that the activation energy may be obtained directly from the slope of a linear fit to $\ln(\sigma)$ versus $1/T$. 

\begin{figure*}[htb] 
    \centering 
    \includegraphics[width=\columnwidth]{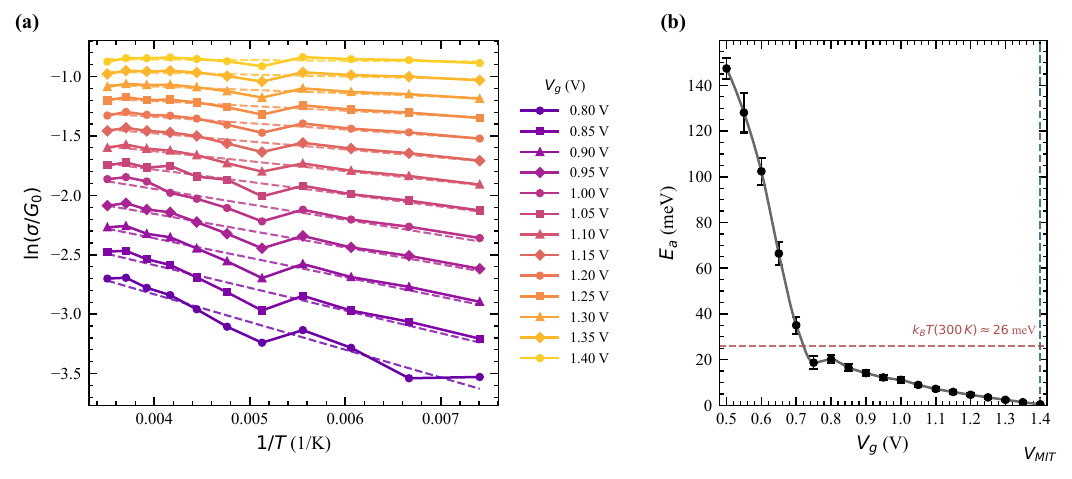} 
    \caption{\textbf{Arrhenius analysis of the insulating-side transport.} 
    (a) Conductivity data plotted as $\ln(\sigma)$ versus $1/T$ for gate voltages between $0.8~\mathrm{V}\leq V_{\mathrm g}\leq1.4~\mathrm{V}$. Solid symbols show the measured data and dashed lines show linear fits performed over the temperature range $135~\mathrm{K}\leq T\leq285~\mathrm{K}$. This gate-voltage range was used in the main text to identify the transition through the $E_{\mathrm a}\rightarrow0$ condition. (b) Corresponding activation energy as a function of gate voltage over the full reliable measurement range of the primary device. The dashed horizontal line indicates the room-temperature thermal energy, $k_{\mathrm B}T\approx26~\mathrm{meV}$ at $300~\mathrm{K}$.}
    \label{fig:S3} 
\end{figure*}

The analysis used in the main text to identify the transition through the $E_{\mathrm a}\rightarrow0$ condition was restricted to gate voltages in the range $0.8~\mathrm{V}\leq V_{\mathrm g}\leq1.4~\mathrm{V}$ and conductivity data acquired over the temperature range $135~\mathrm{K}\leq T\leq285~\mathrm{K}$. Although transport remains insulating at lower gate voltages, the conductivity becomes progressively smaller and the extracted activation energies become increasingly sensitive to experimental noise, leading to larger fitting uncertainties. Reliable activation energies could nevertheless be obtained down to $V_{\mathrm g}=0.5~\mathrm{V}$, as shown in Fig.~\ref{fig:S3}(b). At temperatures below approximately 135 K, systematic deviations from simple Arrhenius behavior become increasingly apparent and were therefore excluded from the fitting procedure.

Figure~\ref{fig:S3}(a) shows the Arrhenius fits used to extract the activation energies reported in Fig.~4(a) of the main text. Activation energies were obtained from the slopes of the corresponding linear regressions, while the associated uncertainties were determined from the fitting procedure and are reflected by the error bars shown in Fig.~4(a) of the main text.

Figure~\ref{fig:S3}(b) extends the activation-energy analysis to the full reliable gate-voltage range of the primary device. While only the range $0.8~\mathrm{V}\leq V_{\mathrm g}\leq1.4~\mathrm{V}$ was used in the main text to identify the transition through the $E_{\mathrm a}\rightarrow0$ condition, activation energies continue to increase systematically at lower gate voltages. The extracted activation barriers reach approximately $147~\mathrm{meV}$ at the lowest reliable gate voltages investigated. For comparison, the horizontal dashed line indicates the room-temperature thermal energy, $k_{\mathrm B}T\approx26~\mathrm{meV}$ at $300~\mathrm{K}$. The experimentally extracted activation barriers therefore substantially exceed room-temperature thermal energies over a significant portion of the insulating regime.

\clearpage
\section{Carrier-density calibration} 

The critical carrier density reported in the main text is obtained by converting the independently determined transition voltage $V_{\mathrm{MIT}}$ into a carrier density using Hall-effect measurements. This procedure relies on an experimentally determined gate capacitance extracted directly from the Hall density measured in the high-density regime. 

\begin{figure*}[htb] 
    \centering 
    \includegraphics[width=\columnwidth]{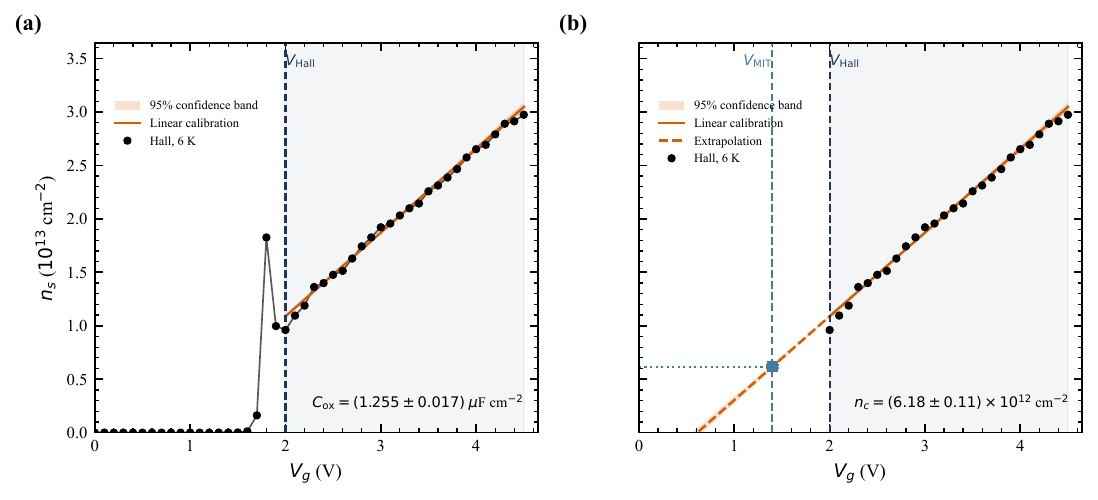} 
    \caption{ \textbf{Carrier-density calibration from Hall measurements.} (a) Hall carrier density as a function of gate voltage measured at 6 K. The linear Hall-density regime above $V_{\mathrm{Hall}}$ is used to extract the effective gate capacitance. Shaded regions indicate pointwise 95\% confidence bands. (b) Extrapolation of the calibrated Hall-density relation to the independently determined transition voltage $V_{\mathrm{MIT}}$. }
    \label{fig:S4} 
\end{figure*}

Figure~\ref{fig:S4}(a) shows the Hall carrier density $n_{\mathrm{s}}$ as a function of gate voltage for the primary device. Above $V_{\mathrm{Hall}}$, the Hall density exhibits an approximately linear dependence on gate voltage, consistent with electrostatic charge accumulation through a constant geometric gate capacitance. Reliable Hall densities can only be extracted above $V_{\mathrm{Hall}}$. Although a system-spanning conducting path already exists at $V_{\mathrm{MIT}}$, the conducting network remains relatively sparse near the percolation threshold and does not necessarily support a measurable transverse Hall signal. As the carrier density increases, the conducting puddles expand, the delocalized carrier population grows, and the Hall voltage becomes sufficiently large for robust carrier-density extraction. Consequently, $V_{\mathrm{Hall}}$ is not a fundamental material parameter but an operational quantity. Its value depends on the strength of the Hall signal relative to the measurement noise floor. Since the Hall voltage scales with carrier mobility and magnetic field, improvements in either quantity can shift $V_{\mathrm{Hall}}$ closer to $V_{\mathrm{MIT}}$. The linear Hall-density regime above $V_{\mathrm{Hall}}$ nevertheless provides a direct experimental calibration of the gate capacitance and carrier density.

A linear fit to the Hall data yields an experimental gate capacitance through 

\begin{equation} 
    n_{\mathrm{s}} = n_{\mathrm{s,ref}} + \frac{C_{\mathrm{ox}}}{q} \left( V_{\mathrm{g}} - V_{\mathrm{ref}} \right), 
\end{equation} 

where $C_{\mathrm{ox}}$ is the effective gate capacitance and $q$ is the elementary charge. 

A linear fit to the Hall density yields an experimental gate capacitance of $C_{\mathrm{ox}}=(1.255\pm0.017)\,\mu\mathrm{F\,cm^{-2}}$. This value is in reasonable agreement with the geometric capacitance, $C_{\mathrm{geom}} = 1.369\,\mu\mathrm{F\,cm^{-2}}$, calculated from the dielectric stack thicknesses and dielectric constants, providing additional validation of the Hall-density calibration procedure. 

Figure~\ref{fig:S4}(b) illustrates the conversion of $V_{\mathrm{MIT}}$ into a critical carrier density. Because reliable Hall densities can only be extracted above $V_{\mathrm{Hall}}$, the experimentally determined linear $n_{\mathrm{s}}$-$V_{\mathrm{g}}$ relation was extrapolated to lower gate voltages. The critical density $n_{\mathrm{c}}$ was then obtained by evaluating this relation at the independently determined transition voltage $V_{\mathrm{MIT}}=1.4~\mathrm{V}$, yielding \[ n_{\mathrm c} = (6.18\pm0.11)\times10^{12} ~\mathrm{cm}^{-2}. \]

This procedure assumes that the capacitance extracted from the Hall-accessible regime remains approximately constant over the extrapolation interval between $V_{\mathrm{Hall}}$ and $V_{\mathrm{MIT}}$. Because the extrapolation spans only a limited gate-voltage range, the statistical uncertainty propagated from the linear Hall-density calibration remains modest. Quoted uncertainties are one-standard-error estimates obtained from the unweighted linear Hall-density calibration. The uncertainty in $n_{\mathrm c}$ includes the slope-intercept covariance of the linear regression, with $V_{\mathrm{MIT}}$ held fixed. Uncertainties associated with individual Hall-coefficient extractions, uncertainty in $V_{\mathrm{MIT}}$, and systematic calibration errors are not included. Shaded regions in Fig.~\ref{fig:S4} indicate pointwise 95\% confidence bands.

\clearpage
\section{Additional scaling analyses} 

To evaluate the robustness of the conductivity scaling presented in the main text, additional scaling analyses were performed over a range of temperatures and assumed transition voltages. For all conductivity scaling analyses, fitting was restricted to the metallic side of the transition and to the density interval $0<n_{\mathrm s}-n_{\mathrm c}<10^{13}~\mathrm{cm}^{-2}$. This choice focuses the analysis on the transport regime closest to the percolation threshold, where the percolation scaling relation is expected to be most reliable. The critical density was treated as a free fitting parameter and optimized by maximizing the coefficient of determination ($R^2$) within the specified fitting window, with candidate densities scanned over the interval $0.90\,n_{\mathrm{s,min}} \leq n_{\mathrm c} \leq 0.99\,n_{\mathrm{s,max}}$.

\begin{figure*}[htb] 
    \centering 
    \includegraphics[width=\columnwidth]{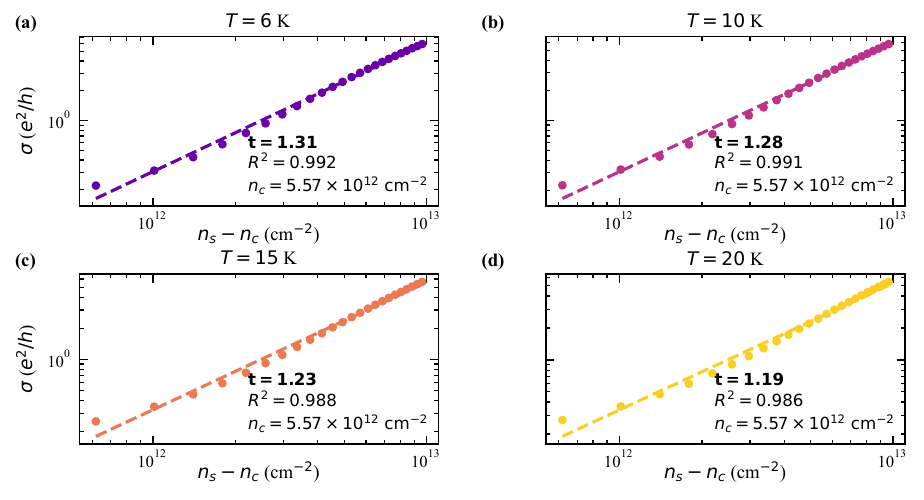} 
    \caption{ \textbf{Conductivity scaling at the lowest measured temperatures.} 
    Conductivity plotted as a function of $(n_{\mathrm s}-n_{\mathrm c})$ for (a) 6 K, (b) 10 K, (c) 15 K, and (d) 20 K using the independently determined transition voltage $V_{\mathrm{MIT}}=1.4~\mathrm{V}$. Dashed lines show fits to the percolation scaling relation. }
    \label{fig:S5} 
\end{figure*}

Figures~\ref{fig:S5}(a--d) show conductivity scaling analyses performed at the four lowest measured temperatures using the same transition voltage, $V_{\mathrm{MIT}}=1.4~\mathrm{V}$, independently determined from the activation-energy analysis. In each case, excellent agreement is obtained with the percolation scaling relation

\begin{equation} 
    \sigma \propto (n_{\mathrm s}-n_{\mathrm c})^t . 
\end{equation} 

The corresponding percolation exponents are $t=1.31$, 1.28, 1.23, and 1.19 for 6, 10, 15, and 20 K, respectively, while the coefficients of determination range between 0.986 and 0.993. While the extracted exponent exhibits a systematic temperature dependence, values obtained at the lowest temperatures remain close to the theoretical two-dimensional percolation value of $t=4/3$. Similar temperature evolution of percolation-fit parameters has been reported in previous conductivity-scaling studies of the two-dimensional metal-insulator transition \cite{das_sarma_two-dimensional_2005}. Because the percolation scaling relation is expected to be most rigorously applicable in the zero-temperature limit, the 6 K result provides the most direct comparison with theory. The extracted exponent of $t=1.31$ at 6 K is in close agreement with the expected two-dimensional percolation value, while maintaining excellent fit quality ($R^2=0.993$). The persistence of good scaling behavior across the lowest measured temperatures demonstrates that the conclusions drawn from the 6 K scaling analysis presented in the main text are robust.

The scaling fits yield critical densities near $n_{\mathrm c}\approx5.57\times10^{12}\,\mathrm{cm}^{-2}$. However, the extracted values of $t$ and $n_{\mathrm c}$ exhibit some dependence on the fitting window and on the allowed range of candidate critical densities. While modest improvements in $R^2$ can be obtained by extending the fitting range, the corresponding changes in the extracted percolation parameters become substantial. For this reason, the conductivity scaling analysis is used primarily as an independent validation of the transition voltage obtained from the $E_{\mathrm a}\rightarrow0$ condition, rather than as a standalone determination of the critical density.

\begin{figure*}[htb] 
    \centering 
    \includegraphics[width=0.65\columnwidth]{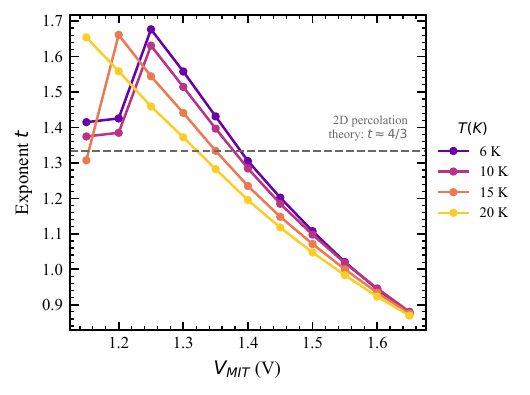}
    \caption{ \textbf{Dependence of the extracted percolation exponent on the assumed transition voltage.} 
    Extracted percolation exponent $t$ as a function of the transition voltage used in the conductivity scaling analysis. The horizontal dashed line indicates the theoretical two-dimensional percolation value $t=4/3$. }
    \label{fig:S6} 
\end{figure*}

Figure~\ref{fig:S6} provides a direct test of the transition voltage identified through the activation-energy analysis. When the conductivity scaling analysis is performed using $V_{\mathrm{MIT}}=1.4~\mathrm{V}$, the extracted exponent is closest to the theoretical two-dimensional percolation value of $t=4/3$. Both larger and smaller assumed transition voltages systematically drive the extracted exponent away from the expected percolation value. This behavior is significant because the transition voltage was determined independently from the $E_{\mathrm a}\rightarrow0$ condition rather than optimized to improve the conductivity scaling. The agreement obtained at $V_{\mathrm{MIT}}=1.4~\mathrm{V}$ therefore provides an independent validation of the transition identified through the activation-energy analysis and supports its interpretation as the disorder-driven percolation threshold.

\clearpage
\section{Reproducibility across devices} 

To verify that the Hall-density collapse and activation-energy signatures reported in the main text are not specific to a single device, equivalent measurements were performed on three additional monolayer MoS$_2$/HfO$_2$ transistors. Together with the primary device discussed in the main text, the investigated set comprises four independent devices spanning different gate geometries and dielectric thicknesses.

\begin{figure*}[htb] 
    \centering 
    \includegraphics[width=\textwidth]{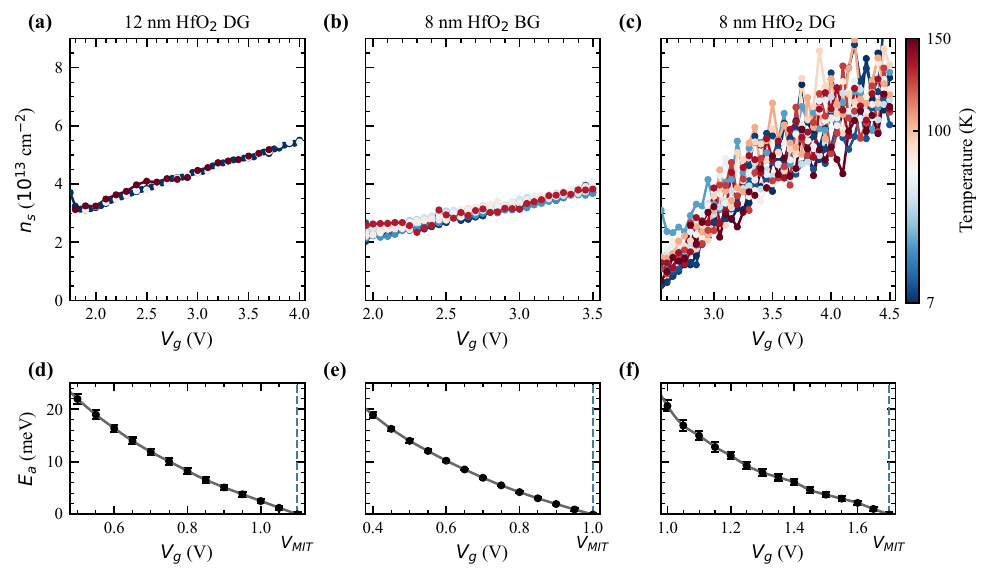} 
    \caption{ \textbf{Reproducibility of the Hall-density collapse and activation-energy signatures across additional devices.} 
    (a--c) Hall density as a function of gate voltage measured for three additional monolayer MoS$_2$ devices spanning different gate geometries and dielectric thicknesses. In each case, the Hall density exhibits a temperature-independent collapse when plotted as a function of absolute gate voltage. Measurements are shown up to $150~\mathrm{K}$ for the secondary devices. (d--f) Activation energy extracted from Arrhenius analysis for the same devices. The activation energy decreases systematically with increasing gate voltage and approaches zero near the indicated transition voltage in each case, which remains below the onset of the Hall-accessible regime. }
    \label{fig:figS7} 
\end{figure*}

Figures~\ref{fig:figS7}(a--c) present the Hall density as a function of gate voltage for the three additional devices. In each case, a temperature-independent Hall-density collapse is observed when the data are plotted as a function of absolute gate voltage, consistent with the behavior reported for the primary device in Fig.~2(a) of the main text. For the secondary devices, Hall densities are shown only up to approximately $150~\mathrm{K}$. Despite this reduced temperature range, the collapse remains clearly observable across all three devices, providing strong evidence that the Hall-collapse signature is reproducible across multiple device configurations. At higher temperatures, the Hall voltage becomes increasingly difficult to resolve at the magnetic fields employed in this study, resulting in insufficient signal-to-noise ratio for reliable carrier-density extraction. This challenge also highlights the difficulty of obtaining reliable Hall-density measurements at elevated temperatures in these devices. In the primary device, the Hall signal remained sufficiently robust to permit carrier-density extraction over a substantially wider temperature range. The dual-gated 8 nm HfO$_2$ device exhibits increased scatter in the extracted Hall densities due to reduced signal-to-noise ratio, although the overall collapse remains evident.

Figures~\ref{fig:figS7}(d--f) show the corresponding activation energies extracted from Arrhenius analysis. In all three devices, the activation energy decreases systematically with increasing gate voltage and approaches zero at a well-defined transition voltage, consistent with the behavior reported in Fig.~4(a) of the main text. In each case, the transition voltage identified through the $E_{\mathrm a}\rightarrow0$ condition remains below the gate-voltage range where reliable Hall densities can be extracted, consistent with the voltage hierarchy established for the primary device. The observation of a vanishing activation energy across multiple devices further supports the identification of the percolation threshold through the $E_{\mathrm a}\rightarrow0$ condition.

Taken together, the four devices investigated in this work consistently exhibit both of the key signatures used to identify the percolation threshold: (i) a temperature-independent Hall-density collapse versus absolute gate voltage and (ii) the vanishing of activation energy at the transition. These observations demonstrate that the percolative conduction onset identified in the main text is reproducible across four independent devices spanning multiple gate geometries and dielectric thicknesses, and is therefore unlikely to arise from a device-specific anomaly.

\clearpage
\section{Disorder energy scale and temperature relevance} 

\begin{table}[htb] 
    \caption{\textbf{Maximum reliable activation barriers extracted from the investigated devices.} 
    The reported activation barrier corresponds to the largest reliable value obtained from Arrhenius analysis for each device. Also shown are the corresponding values of $E_{\mathrm{a,max}}/k_{\mathrm B}T(300~\mathrm{K})$ together with the transition voltages $V_{\mathrm{MIT}}$ and $V_{\mathrm{Hall}}$. Room-temperature thermal energy corresponds to $k_{\mathrm B}T(300~\mathrm{K}) = 25.85~\mathrm{meV}$.}
    \label{tab:S1} 
    \begin{ruledtabular} 
    \begin{tabular}{lcccc} 
    Device & 
    $V_{\mathrm{MIT}}$ ($\mathrm{V}$) & 
    $V_{\mathrm{Hall}}$ ($\mathrm{V}$) &
    $E_{\mathrm{a,max}}$ ($\mathrm{meV}$) & 
    $E_{\mathrm{a,max}}/k_{\mathrm B}T(300~\mathrm{K})$ \\ 
    \hline 
    12 nm HfO$_2$ BG device (primary) 
    & $\sim1.4$ & $\sim2.0$ & 147.4 & 5.7 \\ 
    
    12 nm HfO$_2$ DG device 
    & $\sim1.1$ & $\sim1.8$ & 57.4 & 2.2 \\ 
    
    8 nm HfO$_2$ BG device 
    & $\sim1.0$ & $\sim2.0$ & 67.6 & 2.6 \\ 
    
    8 nm HfO$_2$ DG device
    & $\sim1.6$ & $\sim2.5$ & 121.0 & 4.7 \\ 
    \end{tabular}
    \end{ruledtabular} 
\end{table}

The activation-energy analysis presented in the main text and extended in Fig.~\ref{fig:S3}(b) provides information not only about the transition voltage through the $E_{\mathrm a}\rightarrow0$ condition, but also about the characteristic energy scale associated with the disorder landscape. Table~\ref{tab:S1} summarizes the maximum reliable activation barriers extracted from all investigated devices together with the corresponding values of $V_{\mathrm{MIT}}$ and $V_{\mathrm{Hall}}$.

Across the investigated devices, the maximum reliable activation barriers range from approximately $57$--$147~\mathrm{meV}$. For comparison, room-temperature thermal energy corresponds to $k_{\mathrm B}T\approx26~\mathrm{meV}$ at $300~\mathrm{K}$. Even the smallest experimentally extracted activation barrier exceeds room-temperature thermal energies by more than a factor of two, while the largest exceeds it by nearly a factor of six. 

Although the activation energy represents an effective transport barrier rather than the full disorder-potential amplitude, the extracted values are of the same order of magnitude as oxide-induced electrostatic potential fluctuations predicted by first-principles calculations of TMD/oxide interfaces~\cite{van_troeye_impact_2025}. In a spatially inhomogeneous disorder landscape, local potential fluctuations are expected to span a distribution of larger and smaller values, whereas the experimentally extracted activation energy reflects a single effective transport barrier averaged over the entire device. The similarity of these energy scales therefore provides qualitative support for a common disorder-driven origin. These observations suggest that the disorder landscape responsible for percolative conduction onset remains relevant over a significant fraction of the operating bias range and is not restricted to cryogenic transport.

\bibliography{refs}